\documentclass[12pt]{article}
\usepackage{mathptmx}
\usepackage{amsmath}
\usepackage{amssymb}
\usepackage{amstext}
\usepackage{color}
\usepackage{rotating}
\usepackage{graphicx}
\usepackage{url}
\usepackage{float}
\usepackage{sidecap}
\usepackage{subfig}
\usepackage[breaklinks,colorlinks,linktocpage,final]{hyperref}
\hypersetup{urlcolor=blue}
\usepackage{listings}
\newcommand{\doi}[1]{\href{http://dx.doi.org/#1}{DOI:#1}}
\newcommand{\arxiv}[1]{\href{http://arxiv.org/abs/#1}{arXiv:#1}}

\newcommand{\bq}{\begin{equation}}
\newcommand{\eq}{\end{equation}}

\newcommand{\flop}{\mbox{flop}}
\newcommand{\flops}{\mbox{flops}}

\newcommand{\GBS}{\mbox{GB/s}}

\newcommand{\GFS}{\mbox{GF/s}}

\newcommand{\GHZ}{\mbox{GHz}}

\newcommand{\bytes}{\mbox{bytes}}
\newcommand{\byte}{\mbox{byte}}
\newcommand{\GB}{\mbox{GB}}
\newcommand{\KB}{\mbox{kB}}

\newcommand{\eos}{~.}
\newcommand{\cma}{~,}
\newcommand{\nvidia}{nVidia}
\begin{document}\pagestyle{empty}
\begin{center}\LARGE
Sparse matrix-vector multiplication
on GPGPU clusters:\\A new storage format and a scalable implementation
\\[0.5cm]\large
Moritz Kreutzer$^1$, Georg Hager$^1$, Gerhard Wellein$^{1}$, Holger Fehske$^2$,\\Achim Basermann$^3$, Alan R. Bishop$^4$\\[1mm]\small\itshape
${}^1$Erlangen Regional Computing Center, Erlangen, Germany\\
${}^2$Ernst-Moritz-Arndt University of Greifswald, Greifswald, Germany\\
${}^3$German Aerospace Center (DLR), Simulation and Software Technology, Cologne, Germany\\
${}^4$Theory, Simulation and Computation Directorate, Los Alamos National Laboratory, Los Alamos, NM, USA\\
\end{center}\small
\noindent\textbf{Abstract:}
Sparse matrix-vector multiplication (spMVM) is the dominant
operation in many sparse solvers. We investigate performance
properties of spMVM with matrices of various sparsity patterns on
the \nvidia{} ``Fermi'' class of GPGPUs. A new ``padded
jagged diagonals storage'' (pJDS) format is proposed which may
substantially reduce the memory overhead intrinsic to the
widespread ELLPACK-R scheme. In our test scenarios the pJDS format
cuts the overall spMVM memory footprint on the GPGPU by up to 70\%,
and achieves 95\% to 130\% of the
ELLPACK-R performance.  Using a suitable performance model we
identify performance bottlenecks on the node level that invalidate
some types of matrix structures for efficient multi-GPGPU
parallelization. For appropriate sparsity patterns we extend
previous work on distributed-memory parallel spMVM to demonstrate a
scalable hybrid MPI-GPGPU code, achieving efficient overlap of
communication and computation.
\normalsize

\section{Introduction and related work}

\subsection{Sparse matrix-vector multiplication}

The solution of large eigenvalue problems or extremely sparse systems
of linear equations is a central part of many numerical algorithms in
science and engineering. Sparse matrix-vector multiplication (spMVM)
is often the dominating component in such solvers, and may easily
consume most of the total runtime. General-purpose computation on
graphics processing units (GPGPUs) is an attractive option for this
operation due to the large memory bandwidth available to high-end
graphics chips and their inherent massive parallelism.
Implementations of spMVM on GPGPUs have been a field of active
research in recent years~\cite{bg09,spauto,ellr11}, and several
storage formats have been proposed. Out of those, the ELLPACK-R format
\cite{ellr11} has gained widespread acceptance. However, although
there is a long history of distributed-memory parallel spMVM codes
(see~\cite{lspp11} and references therein), there is to our knowledge
no efficiency or feasibility analysis of multi-GPU spMVM.

This work has two goals: It provides an alternative sparse MVM storage 
format that has a significantly smaller memory footprint than ELLPACK(-R)
but provides better performance in most cases on modern \nvidia\ GPGPUs. 
Furthermore, it extends previous work on distributed-memory spMVM
for general matrices to multiple GPGPUs.

\subsection{Testbed}

The \nvidia{} ``Fermi'' class of GPGPU-based accelerators (Tesla C/M20X0) used
for the benchmarks implement the ``GF100'' architecture and comprise
14 \emph{streaming multiprocessors} (MPs), each with 32 in-order
cores. One core can execute one single-precision (SP) multiplication
and one addition per cycle, which leads to an overall peak performance
of 896\,\flops\ per cycle on the whole chip at clock frequencies above
1\,\GHZ. At double precision (DP) the theoretical peak performance is
halved. The boards are currently available with device memory sizes
of 3 (C2050) or 6\,\GB\ (C2070), and feature
deactivatable ECC protection.  In streaming benchmarks the device
memory delivers about 91\,\GBS\ sustained with ECC enabled ($120$~GB/s
w/o ECC)~\cite{gpulbm}. All cores share a
768\,\KB\ L2 cache, whose detailed specifications are undisclosed.

The cores in an MP are driven in a single instruction multiple data
(SIMD) manner (also termed SIMT model, where ``T'' stands for
``threads'').  All threads running on an MP are controlled by
a simple instruction scheduler that can switch quickly between chunks
of threads called \emph{warps}, in order to hide latencies. 
A warp (or a subset of it) is the actual SIMD unit on this device, and 
it is essential that consecutive threads in a warp access consecutive
memory locations (this is called \emph{coalescing}). Although 
still important, coalescing constraints have been somewhat relaxed
with the GF100 architecture due to its L2 cache, which was not present 
on earlier models.

The parallel runs have been conducted on the \emph{Dirac}%
\footnote{http://www.nersc.gov/users/computational-systems/dirac/} GPGPU 
cluster at the National Energy Research Scientific Computing Center (NERSC)
in Berkeley. This cluster features 50 GPU nodes, of which 44 
contain one \nvidia{} Tesla C2050 card with 3\,\GB\ of 
device memory.

\subsection{Test matrices}\label{sect:matrices}

\textbf{HMEp}: This matrix originates from the quantum-mechanical description
(using the so-called Holstein-Hubbard model)
of a one-dimensional solid with six lattice sites, populated with
six electrons coupled to 15 phonons (quantized lattice vibrations).
The resulting matrix of
dimension $6.2\times 10^6$ is very sparse, with approximately
15 non-zeros per row. It also contains contiguous off-diagonals
of length 15,000.

\noindent\textbf{sAMG}:
This matrix was generated by the adaptive multigrid code sAMG
(see~\cite{AMG,sAMG}, and references therein) for the irregular discretization
of a Poisson problem on a car geometry. Its matrix dimension is
$3.4\times10^6$ with an average of $N_\mathrm{nzr}\approx 7$ entries
per row.

\noindent\textbf{DLR1}:
This matrix comes from an adjoint problem computation 
(turbulent transonic flow over a wing) with 
the TAU CFD system of the German Aerospace Center (DLR).  
TAU performs complex flow simulations
on unstructured hybrid grids. The associated grid had 46,417
points (6 unknowns in each point), and the resulting matrix is 
nonsymmetric with a dimension of $2.8 \times 10^5$ and an average of
$N_\mathrm{nzr} \approx 144$ entries per row.

\noindent\textbf{DLR2}:
This matrix stems from a linear problem for an aerodynamic
gradients calculation.  A transonic inviscid flow over a
wing was simulated with TAU.  The associated grid had
108,396 points, and the matrix is nonsymmetric with a  dimension
of $5.4 \times 10^5$ and an average of $N_\mathrm{nzr} \approx 315$ entries
per row. It consists entirely of dense $5\times 5$ subblocks.

\noindent\textbf{UHBR}:
The last matrix originates from
aeroelastic stability investigations of an ultra-high bypass ratio
(UHBR) turbine fan with a
linearized Navier-Stokes solver~\cite{UHBR}. This solver is part of
the parallel simulation system TRACE (Turbo-Machinery Research
Aerodynamic Computational Environment) which was developed by DLR's
Institute for Propulsion Technology. Its matrix dimension is
$4.5\times10^6$ with an average of $N_\mathrm{nzr}\approx 123$ entries
per row.

\section{GPGPU spMVM}

\subsection{Introducing the padded JDS formats}\label{sect:storage}

\begin{figure}[tbp]
\centering
\includegraphics*[width=0.85\textwidth]{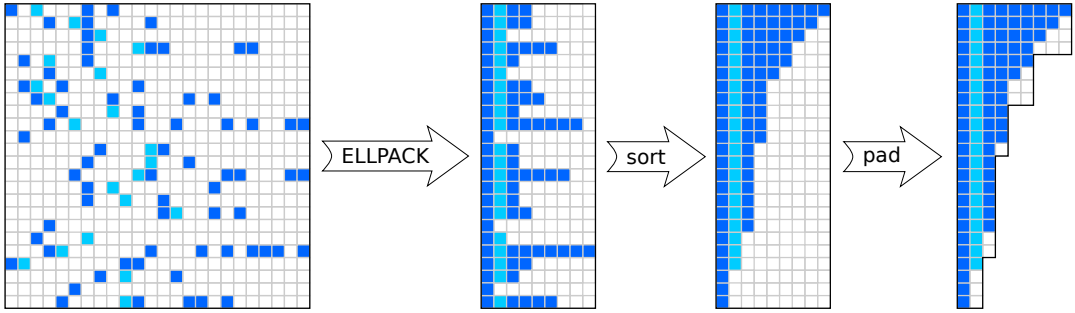}
\caption{Derivation of the pJDS format from a sparse matrix. 
  In the pJDS format a blocking size of $b_r=4$ is 
  used.\label{fig:pJDSformat}}
\end{figure}

ELLPACK-R \cite{ellr11} is a variant of the original
ELLPACK storage format~\cite{bg09,gky79} and sets today the standard
for implementing spMVM operations on GPGPUs. ELLPACK(-R) should be
used if no regular substructures such as off-diagonals or dense blocks 
can be exploited. The idea is to compress
the rows by shifting all non-zero entries to the left (first
step in Fig.~\ref{fig:pJDSformat}) and storing the resulting $N \times
N^\textrm{max}_\mathrm{nzr}$ rectangular matrix\footnote{Typically
  the matrix dimension $N$ must also be padded to a multiple of
  the warp size.} column-by-column consecutively in main memory,
where $N^\textrm{max}_\mathrm{nzr}$ is the maximum number of non-zeros
per row. Thus,
in contrast to CPU storage formats, ELLPACK contains zero entries
(white boxes in Fig.~\ref{fig:pJDSformat}). 
Thread parallelization of the spMVM is 
row-wise by assigning consecutive rows to the threads of a
block (i.e., outer loop iterations in
Listing~\ref{listing:ellpack} are mapped to threads in a round-robin way). 
\lstset{caption={The standard ELLPACK-R spMVM kernel}\label{listing:ellpack}}
\begin{lstlisting}
 for(i=0; i < N; ++i)
     for(j=0; j < rowmax[i]; ++j)
         c[i] += val[j*N + i] * rhs[ col_idx[j*N + i] ];
\end{lstlisting}
The increased memory footprint of the ELLPACK format
ensures load coalescing within thread warps for access to the matrix
entries (\verb.val[].) and the index array (\verb.col_idx[].), which
points to the right hand side (RHS) vector elements (\verb.rhs[].).
While data alignment became of minor importance with the latest \nvidia{}
GPGPU generations, load coalescing is still a must for attaining
reasonable data transfer rates. In the original ELLPACK scheme the
threads were still loading and operating on the zero matrix
entries, wasting memory bandwidth and compute resources.

\begin{SCfigure}[0.95]
\begin{tabular}{lll}
\includegraphics*[width=0.15\textwidth]{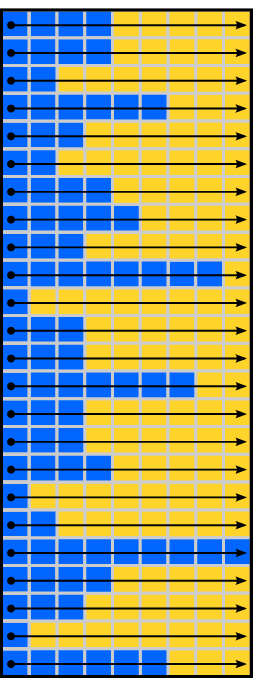}~~~ &
\includegraphics*[width=0.15\textwidth]{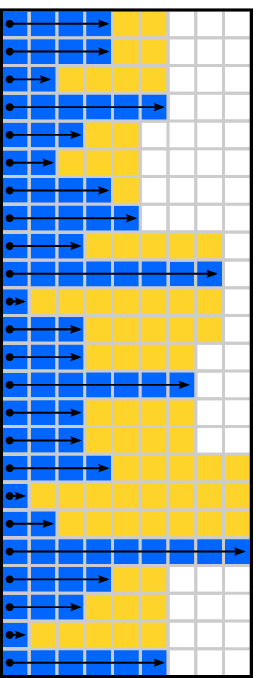}~~~ &
\includegraphics*[width=0.15\textwidth]{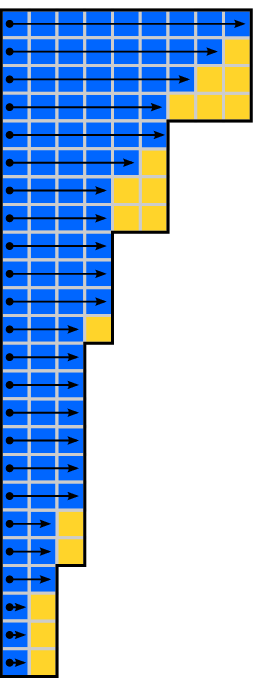} \\
(a) ELLPACK & (b) ELLPACK-R & (c) pJDS
\end{tabular}
\caption{Scheduling patterns and required storage size for the
  different matrix formats assuming a four thread warp. Arrows
  indicate computation and data accesses executed by the threads. White
  boxes show redundant data storage, and light boxes indicate redundant
  data storage and useless hardware reservation.\label{fig:sched}}
\end{SCfigure}

The ELLPACK-R scheme uses the same storage format, but threads
only execute non-zero contributions (the number of non-zeros per row is
stored in \verb.rowmax[].), avoiding redundant data transfers. However,
all threads of a warp occupy on-chip resources until the
thread executing the longest row has finished.
Figures~\ref{fig:sched}a and b compare the overhead of the
ELLPACK(-R) schemes assuming a warp size of four threads. 
ELLPACK-R reduces computation and data accesses to the possible
minimum (arrows in Fig.~\ref{fig:sched}b).
However, the imbalanced row lengths impose reservation of unused hardware
units (light boxes). Moreover the redundant storage
(indicated by white and light boxes) stays the same.

A simple idea, derived from the 
Jagged Diagonals Storage (JDS) format used for vector computers, 
can drive the matrix format towards better
utilization of compute resources and storage. First the rows
of the ELLPACK scheme are sorted according to the number of
non-zeros, starting with the longest row (``sort'' step in
Fig.~\ref{fig:pJDSformat}). Then, blocks of $b_r$ consecutive rows
(where $b_r$ should be the warp size) are padded to the
longest row within the block (``pad'' step in
Fig.~\ref{fig:pJDSformat}). We call the result
``padded Jagged Diagonals Storage'' (pJDS). This 
maintains load coalescing while most of the zero
entries can be eliminated. Since the columns typically have
different lengths, a (small) array \verb.col_start[]. of size
($N^\textrm{max}_\mathrm{nzr} \times 4 $\,\byte) is required to store the
starting offset of each column. The pJDS kernel is shown in
listing~\ref{listing:pJDS}.
\lstset{caption={The spMVM kernel of the pJDS format}\label{listing:pJDS}}
\begin{lstlisting}
 for(i=0; i < N; ++i)
     for(j=0; j < rowmax[i]; ++j){
         col_offset = col_start[j];
         c[i] += val[col_offset + i] * rhs[ col_idx[col_offset + i] ];
     }
\end{lstlisting}
It maintains the
structure and simplicity of the ELLPACK-R kernel but provides
potential for (substantial) data reduction and better hardware
utilization. The main drawback of the format is that the spMVM
operation needs to be performed in a permuted basis. However, for most
iterative spMVM algorithms such as Krylov subspace methods,
permutation of the indices needs to be done only before the start and
after the end of the algorithm, while the complete iterative scheme
works on the permuted elements. On the downside, the permutation of the
matrix rows can destroy matrix structures such as off-diagonals or
local dense blocks, leading to a loss of load coalescing or cache
reuse on the RHS vector.

\begin{SCfigure}[0.22]
\includegraphics*[width=.8\textwidth]{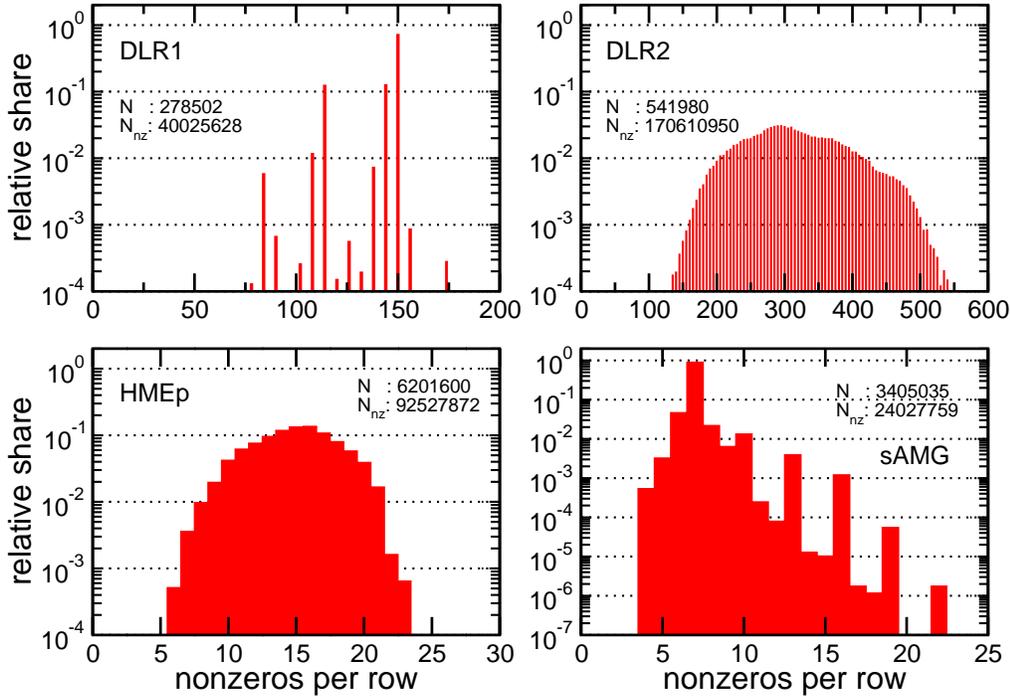}
\caption{Row length distribution histograms of the matrices described 
  in Sect.~\ref{sect:matrices}. The bin size is 1 for all cases.\medskip\label{fig:histogram}}
\end{SCfigure}
The sparsity pattern determines the data reduction
potential of pJDS. If the matrix has a constant row length
(\verb.rowmax[]=.$N^\textrm{max}_\mathrm{nzr}$), ELLPACK and pJDS
both have no storage overhead ($N \times N^\textrm{max}_\mathrm{nzr}$ non-zeros). 
On the other hand, if there is one fully populated row and a single entry in 
all others, the plain
ELLPACK format would store the full matrix, i.e., $N \times N$
elements. In pJDS it is sufficient to hold $b_r \times N +
N - b_r = (b_r+1) \times N - b_r$ entries. At a typical value of
$b_r=32$ one can expect a substantial reduction of the memory
footprint for matrices with a wide variation in row lengths.

The row length histograms (Fig.~\ref{fig:histogram}) for DLR1/2, sAMG, and HMEp
show that there is plenty of data reduction potential for those matrices.
DLR1 should benefit least from the pJDS format since it has the
lowest relative width (\verb.max(rowmax[])/min(rowmax[]).$\approx\,2$)
and most of the weight is clustered close to the maximum row length,
i.e., $80\%$ of the rows have a length of $0.8 \times
N^\textrm{max}_\mathrm{nzr}$. In contrast, the longest row of sAMG
is more than four times larger than the smallest one,
and short rows account for most of the weight. The data reduction rates
finally achieved by using pJDS instead of ELLPACK
follow this qualitative discussion and are shown in
Table~\ref{tab:pJDSperf}.
Considering the limited amount and high cost of device memory on
GPGPUs, pJDS  delivers a useful shrink of the memory
footprint for spMVM on GPGPUs; e.g., the DLR2 matrix fits
(in double precision) on an \nvidia\ Fermi C2050 GPGPU only when using the pJDS
format. The overhead of pJDS compared to a minimum
implementation (storing the non-zeros only) is less than 0.01\%
for the matrices considered here (choosing $t_b=32$).
\begin{table}[tb]
\centering
\begin{tabular}{r|r|c|c|c|c}
\multicolumn{2}{c|}{}  &  DLR1  & DLR2   &  HMEp  &  sAMG \\\hline 
\multicolumn{2}{r|}{data reduction [\%]} & 17.5 & 48.0 & 36.0 & 68.4 \\\hline\hline 
     &  ELLPACK-R     & 22.1      & 15.2      & 15.8      & 14.6 \\ 
\begin{rotate}{0}\makebox[0pt][r]{\raisebox{1ex}{SP ECC=0}}\end{rotate}   &  pJDS    & \textbf{27.6} & \textbf{18.7} & \textbf{18.9} & \textbf{19.5} \\\hline 
     &  ELLPACK-R     & \textbf{12.9} & \textbf{9.6}  & \textbf{7.9}   & 7.8  \\ 
\begin{rotate}{0}\makebox[0pt][r]{\raisebox{1ex}{DP ECC=1}}\end{rotate}   &  pJDS    & \textbf{12.9} & 9.5       & 7.5        & \textbf{8.5} \\\hline 
Westmere EP & CRS (DP) & 5.7  & 5.8  & 3.9 & 4.1 \\ 
\end{tabular}
\caption{Data reduction of pJDS in comparison to the ELLPACK format
  and performance (in \GFS, excluding data transfers) 
  of the different storage formats on an
  \nvidia\ Fermi C2070 GPGPU. The best performance for each matrix and
  execution environment (double precision with ECC enabled on vs.
  single precision with ECC disabled) is
  highlighted. 
  The last row shows the performance of a
  dual-socket (12 core) Intel Westmere node. See
  \cite{lspp11} for implementation and hardware details.\label{tab:pJDSperf}}
\end{table}

The improved hardware utilization by pJDS (compare
Figs.~\ref{fig:sched}b and c) is also
reflected in the performance numbers. In most scenarios gains of up to
$30\%$ can be achieved with pJDS, while the largest penalty is
limited to $5\%$. Since the row permutation may destroy regular
substructures, the DLR2 and HMEp matrices do show some
performance drop or only moderate speed-ups due to reduced cache reuse and
load coalescing for the RHS vector. This
problem is more severe on older GPGPU generations
without L2 cache, such as the Tesla C1060. Here it is also
necessary to map the array holding the column starting offsets
(\verb.col_start[].) to the texture cache.

In summary, the pJDS format presents a very attractive alternative to
the ELLPACK-R scheme on modern GPGPUs both in terms of performance and
memory footprint.

\subsection{GPGPU performance model and PCIe transfer impact}\label{sect:perfmod}

Due to the small (or non-existent) data cache on
GPGPUs, the expected speedup compared to a multicore socket is usually smaller
than the ratio of memory bandwidths.
The worst-case code balance of the ELLPACK and pJDS kernels for double precision
is
\bq\label{balance}
B_\mathrm{W}^\mathrm{DP}=\left(\frac{8+4+8\alpha+16/N^\textrm{max}_\mathrm{nzr}}{2}\right)
\frac{\bytes}{\flop} =\left(6+4\alpha+\frac{8}{N^\textrm{max}_\mathrm{nzr}}\right)
\frac{\bytes}{\flop}\eos
\eq
The parameter $1/N^\textrm{max}_\mathrm{nzr}\leq\alpha\leq 1$ 
quantifies the possible re-use of RHS data from cache:
If there is no cache, i.e., if each load to the RHS vector goes to
memory, we have $\alpha=1$. Hence, cache is able to reduce
the balance by some amount. In the ideal case $\alpha=1/N^\textrm{max}_\mathrm{nzr}$
each RHS element has to be loaded only once. This corresponds to the
$\kappa=0$ case in \cite{lspp11}.
Note that $B_\mathrm{W}^\mathrm{DP}$ may change 
from block to block due to different values of $N^\textrm{max}_\mathrm{nzr}$,
and that the \verb.col_start[]. array is assumed to always come from cache.
In the following we assume an average value $N_\mathrm{nzr}$ for the number
of non-zeros per row.

The bandwidth model (\ref{balance}) is only valid for the kernel execution on
the device, and does not include the data transfers required to bring the 
RHS vector to the GPU and the result back to the host. However, one can extend
the model to incorporate those overheads. Since two distinct bandwidths are
involved we now look at the expected wallclock times for the pure spMVM 
($T_\mathrm{MVM}$) and the required data transfer of the RHS and LHS vectors
over the PCIe bus ($T_\mathrm{PCI}$):
\bq
T_\mathrm{MVM}  = \frac{8N}{B_\mathrm{GPU}}\left[N_\mathrm{nzr}
  \left(\alpha+\frac{3}{2}\right)+2\right] \qquad\mbox{and}\qquad
T_\mathrm{PCI} =  \frac{16N}{B_\mathrm{PCI}}\eos
\eq
This shows that a low PCIe bandwidth has less impact on the overall
execution time if $N_\mathrm{nzr}$ is large, hence we can estimate the
range of favorable $N_\mathrm{nzr}$ values: Setting 
$T_\mathrm{MVM}\leq T_\mathrm{PCI}$, i.e., assuming more than 50\% penalty from
the PCIe transfers, we arrive at
\bq
N_\mathrm{nzr}\leq\frac{2(B_\mathrm{GPU}/B_\mathrm{PCI}-1)}{\alpha+3/2}\eos
\eq
In the worst case, $\alpha=1/N_\mathrm{nzr}$ and 
$B_\mathrm{GPU}\gtrsim 20 B_\mathrm{PCI}$ lead to $N_\mathrm{nzr}\leq 25$.
On the other hand, if $\alpha=1$ and $B_\mathrm{GPU}\approx 10 B_\mathrm{PCI}$
we have $N_\mathrm{nzr}\leq 7$. Thus we do not expect a significant
benefit from GPGPU acceleration for the HMeP and sAMG matrices described above,
since those have $N_\mathrm{nzr}\approx 15$ and $7$, respectively.

If we want less than 10\% penalty from PCIe transfers 
($T_\mathrm{MVM}\geq 10 T_\mathrm{PCI}$) we get 
\bq
N_\mathrm{nzr}\geq\frac{20B_\mathrm{GPU}/B_\mathrm{PCI}-2}{\alpha+3/2}\cma
\eq
so at $B_\mathrm{GPU}\approx 10 B_\mathrm{PCI}$ and $\alpha=1$
a value of $N_\mathrm{nzr}\gtrsim 80$ is sufficient. This is certainly
satisfied for all  DLR matrices. In the worst case, i.e., 
at $B_\mathrm{GPU}\approx 20 B_\mathrm{PCI}$ and $\alpha=1/N_\mathrm{nzr}$
one arrives at $N_\mathrm{nzr}\gtrsim 266$; in this case we can expect
a measurable impact of PCIe transfer overhead for the matrices considered here.

\section{Distributed-memory spMVM parallelization}\label{sec:dmpar}

\begin{figure}[tbp]
\centering
\includegraphics*[width=0.85\textwidth]{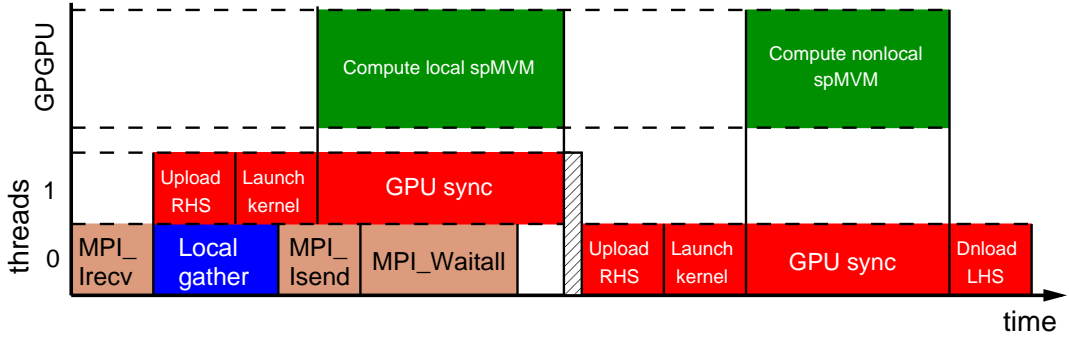}
\caption{Timeline for GPGPU-based spMVM kernel including host data transfers
  with a dedicated host thread for asynchronous MPI communication
  (thread 0).
  The ``local gather'' is the collection of data to be sent to other 
  processes into a contiguous buffer.\label{fig:gpu12}}
\end{figure}
As described in Sect.~\ref{sect:perfmod}, matrices with small $N_\mathrm{nzr}$
are no good candidates for GPGPU acceleration since the required PCIe transfers
for the RHS and LHS vectors dominate the runtime. Although this penalty
is somewhat mitigated by the fact that in some real applications parts of those
vectors may be kept on the device, all data that has to be communicated using
MPI must also cross the PCIe bus to the GPGPU. 
For the HMEp (sAMG) matrix we arrive at a single-GPU performance level of 
3.7 (2.3)~\GFS, which is already below the capability of a typical
dual-socket server node (see Table~\ref{tab:pJDSperf}).
Hence, we restrict the discussion in this section to the DLR1 and UHBR 
matrices. Although they also suffer from PCIe transfers
to some extent (10.9\,\GFS\ vs. 12.9\,\GFS\ for DLR1), there is still 
a substantial advantage over the CPU version.

All runs were performed in double precision and with active ECC 
on the NERSC \emph{Dirac}
cluster.  The ELL\-PACK-R format was used throughout, since the matrix
storage format is of minor importance for the double precision case
(see Table~\ref{tab:pJDSperf}) and for the concepts we want to
demonstrate here. An implementation of the multi-GPGPU code with the pJDS
format and an analysis of its performance implications is ongoing
work and will follow the strategy described in \cite{vecpar02}.

\subsection{Multi-GPGPU spMVM}

The basic design patterns and choices described in \cite{lspp11} for
distributed-memory parallel sparse MVM also apply for the multi-GPGPU case.
We distinguish between \emph{vector mode}, which resembles the
programming style on vector-parallel machines, and \emph{task mode},
which dedicates host resources (threads) to different tasks, i.e., 
communication and computation. In this work we consider three alternatives:

\noindent\emph{Vector mode} without overlap of communication and computation.
  The required communication to distribute the nonlocal RHS elements among
  the processes is separate from the actual spMVM communication, 
  which is performed in a single step.

\noindent\emph{Vector mode} using naive overlap of communication and
  computation by nonblocking MPI. The spMVM must be split into a local
  and a nonlocal part, and the former may be overlapped with MPI.
  Since the result vector must be written twice, there is a slight increase 
  in memory traffic, which adds another $8/N_\mathrm{nzr}$\,\bytes/\flop\ 
  to the code balance (\ref{balance}). Due to the rather large $N_\mathrm{nzr}$
  of the DLR1 and UHBR matrices we expect a performance penalty
  of below 10\%, though. Since most MPI libraries do not support 
  asynchronous nonblocking point-to-point communication, we do not
  expect this variant to have any advantage even over
  vector mode without overlap.

\noindent\emph{Task mode} using a dedicated thread for MPI 
  in order to implement reliably asynchronous communication. Figure~\ref{fig:gpu12}
  shows an event timeline that visualizes the different tasks executed
  on two host threads (or more if there are multiple GPGPUs in a node) and the
  GPGPU. Depending on the ratio of communication to computation time, the 
  possible performance benefit can be at most a factor of two. At 
  strong scaling we expect task mode and vector mode performance to
  converge.

\subsection{Performance results}\label{sec:performance}

\begin{figure}[tbp]
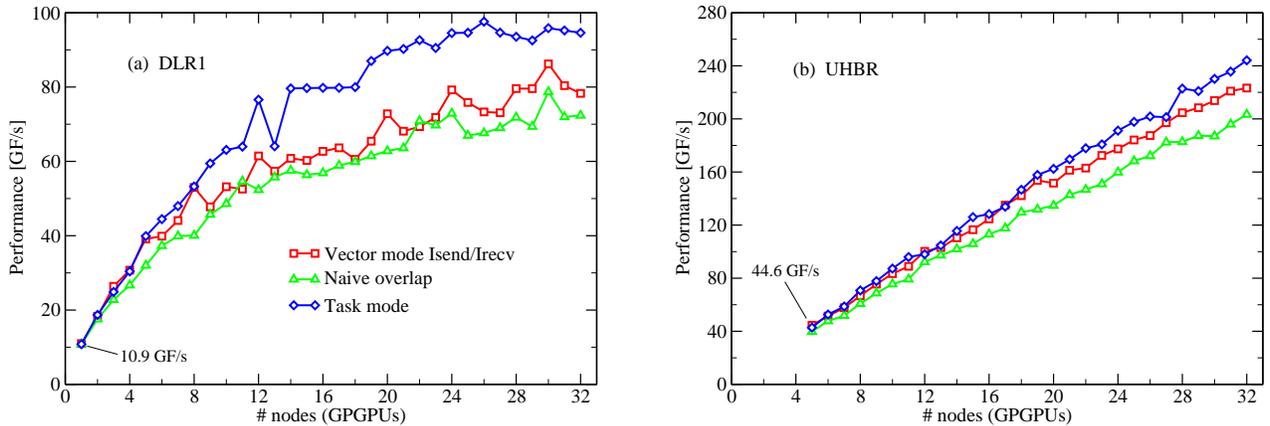

\includegraphics*[width=0.47\textwidth]{dlr1_scaling}
\hfill
\includegraphics*[width=0.47\textwidth]{dlr4_scaling}
\caption{Strong scaling results for DLR1 (left) and UHBR (right) on the Dirac cluster.
  Due to memory restrictions on the C2050 cards it was not possible to 
  run the UHBR case on fewer than five nodes.\label{fig:scaling}}
\end{figure}
Figures~\ref{fig:scaling}a and \ref{fig:scaling}b show strong
scaling results for the DLR1 and UHBR matrices, respectively, on the
Dirac cluster. Task mode achieves better performance
than any of the vector modes in both cases; however, the details
differ considerably:

DLR1 has a rather small dimension of $2.8 \times 10^5$,
so that only 8750 rows (about $1.3\times 10^6$ non-zeros) are left
per GPGPU at 32 nodes. The smallness of the per-GPGPU subproblem
leads to a substantial performance drop, which mainly originates
from the nonlocal part in the naive vector and task mode versions. 
It can, however, be partially
compensated by asynchronous communication. At larger node counts
the performance of all variants starts to converge, as expected.

UHBR has a much larger number of non-zeros at a similar $N_\mathrm{nzr}$
as DLR1, and thus does not show an analogous per-GPGPU performance 
breakdown when scaling up the node count. Scalability is very good
in task mode with a parallel efficiency of 84\% at 32 nodes (about
70\% with naive overlap vector mode).
Since the communication requirements are weaker than for DLR1,
we do not see a similarly large benefit from overlapping
communication at the node counts accessible
on the cluster used.

\section{Conclusions and outlook}

We have introduced a new ``padded JDS'' sparse matrix format, which is
suitable for sparse matrix-vector multiplication on GPGPUs at similar
or better performance levels than the popular ELLPACK-R format,
with a potential for significant memory savings. 

Via suitable performance models we have derived a
condition for the average number of non-zeros per matrix row that
guarantees a useful performance benefit of GPGPU-based spMVM in
comparison to standard server nodes, the main parameter being
the ratio between PCI express bandwidth and GPGPU memory
bandwidth.

Finally we have extended previous work on efficient distributed-memory
hybrid (MPI+OpenMP) spMVM parallelization to the multi-GPGPU case.
Using dedicated host threads for explicitly asynchronous MPI
communication we were able to improve significantly over naive
``vector-like'' approaches and show the potentials and limitations
of this solution. 

Future work will cover more extensive scaling studies on larger GPGPU
clusters, an implementation of the pJDS format in the multi-GPGPU
code, a thorough investigation of the performance degradation with strong
scaling, and the application of our results to a production-grade
eigensolver.

During the preparation of the manuscript it came to our attention that
other research groups have devised sparse matrix formats that share
some features with pJDS, most notably the ``sliced ELLPACK'' and 
``sliced ELLR-T'' formats \cite{monakov10,dziekonski11}. A thorough
comparison of pJDS with those alternative approaches is work in progress.
\bigskip

\textbf{Acknowledgments}\qquad
Discussions with Jan Treibig and Thomas Zeiser are gratefully
acknowledged. We are indebted to Matthias Griessinger for initial
implementations of the GPGPU spMVM kernels, to Gerald Schubert for
providing CPU comparisons, and to K.~St{\"u}ben and H.\,J.~Plum for
providing and supporting the sAMG test case.  This research used the
\emph{Dirac} GPGPU cluster of the National Energy Research Scientific
Computing Center, which is supported by the Office of Science of the
U.S. Department of Energy under Contract No.~DE-AC02-05CH11231. Part
of this work was supported by the competence network for scientific
high performance computing in Bavaria (KONWIHR) via the project
HQS@HPC-II.\vspace*{-5mm}

\small\RaggedRight

\end{document}